\documentclass[aps,prl,twocolumn,showpacs,superscriptaddress]{revtex4}
\usepackage[dvips]{graphicx}
\usepackage{amsmath}
\usepackage{times}

\begin{document}
\author{D. V. Savin}
\author{H.-J. Sommers}
\affiliation{Fachbereich Physik, Universit\"at Duisburg-Essen, 45117 Essen,
Germany}
\author{Y. V. Fyodorov}
\affiliation{School of Mathematical Sciences, University of
Nottingham, Nottingham NG7 2RD, United Kingdom}

\title{Universal statistics of the local Green function in
wave chaotic systems with absorption}
\published{October 25, 2005 in:\quad \texttt{JETP Lett.\,\textbf{82}, 544 (2005)} } %

\begin{abstract}
We establish a general relation between the statistics of the local Green
function for systems with chaotic wave scattering and uniform energy loss
(absorption) and the two-point correlator of its resolvents for the same
system without absorption. Within the random matrix approach this kind of a
fluctuation dissipation relation allows us to derive the explicit analytic
expression for the joint distribution function of the real and imaginary
part of the local Green function for all symmetry classes as well as at an
arbitrary degree of time-reversal symmetry breaking in the system. The
outstanding problem of orthogonal symmetry is further reduced to simple
quadratures. The results can be applied, in particular, to the
experimentally accessible impedance and reflection in a microwave cavity
attached to a single-mode antenna.
\end{abstract}


\maketitle

Statistical fluctuations of physical observables in quantum systems with
underlying chaotic classical dynamics are the subject of a very active field
of research in theoretical and experimental physics. A considerable progress
was underpinned by revealing the apparent universality of the fluctuations
in systems of very diverse microscopic nature, ranging from atomic nuclei
and Rydberg atoms in strong external fields, to complex molecules, quantum
dots, and mesoscopic samples \cite{Haake,Stoeckmann}. This universality
allows one to exploit the random matrix theory (RMT) as a powerful tool for
a theoretical analysis of generic statistical properties of such systems
\cite{Guhr1998}. In many cases the quantity, which is readily obtained
experimentally, is the absorption spectrum for transitions from a given
initial state to highly excited chaotic states at the energy $E$. Most
frequently the absorption spectra look practically continuous due to both
inevitable level broadening and finite experimental resolution. Then the
relevant statistics are the distribution and the correlation functions of
the absorption probability \cite{Fyodorov1998i}. In the simplest situation
of uniform level broadening $\Gamma$, the problem thus amounts to studying
statistical properties of the resolvent (Green function) operator
$\hat{G}(E)\equiv(E+i\Gamma/2-\hat{H})^{-1}$ associated with the random
matrix $\hat{H}$, which replaces the actual chaotic Hamiltonian. In
particular, the imaginary part of diagonal entries of $\hat{G}(E)$ is well
known in solid state physics as the \textit{local density of states} (LDoS)
and in this capacity its statistics enjoyed  many studies
\cite{Fyodorov1998i,Efetov1993,Taniguchi1996,Mirlin2000,Fyodorov2004ii}.

From the experimental point of view the same universality, which makes the
use of the RMT legitimate, provides one with an attractive possibility to
employ simple model systems for analyzing generic statistics of the
fluctuating quantities. Various billiards are nowadays proved to be an ideal
playground for investigating a variety of quantum chaos phenomena; see
Ref.~\cite{Stoeckmann} for a general discussion as well as
\cite{Kuhl2005,Hemmady2005,Kim2005,Lobkis2003,Barthelemy2005a} for the
current problematic. They are usually realized as electromagnetic resonators
in a form of two-dimensional cavities shaped to ensure the chaoticity of
internal scattering and coupled to waveguides (antennas), which are used to
inject microwaves into the system as well as to collect the output. In
particular, it turns out that for a single-mode antenna the local Green
function $G(E)$ (i.e. a diagonal element of $\hat{G}$ taken at the port
position) has the direct physical meaning of the electric impedance of the
cavity which relates linearly voltages and currents at the antenna port; see
\cite{Hemmady2005} for a discussion. In this way not only the imaginary part
but also the real part  of $G$ turn out to be experimentally accessible
quantities whose statistics is, therefore, of considerable interest.
Inevitable energy losses (absorption) must be taken into account properly
when describing the experiments theoretically
\cite{Brouwer1997ii,Savin2003i}.

The majority of the experiments is performed in systems which are
time-reversal invariant (the so-called orthogonal symmetry class of the RMT
characterized by Dyson's symmetry index $\beta=1$). However,
non-perturbative analytical results are available presently only for systems
with no time-reversal symmetry (TRS) ($\beta=2$ or unitary symmetry class)
derived by various methods in \cite{Efetov1993,Fyodorov2004ii}, and for
systems with spin-orbit scattering ($\beta=4$ or symplectic symmetry class)
\cite{Fyodorov2004ii}. An attempt \cite{Taniguchi1996} to provide an
expression for the LDoS distribution for the $\beta=1$ case can not be
considered as particularly successful, as the final expression was given in
a form of an intractable fivefold integral. Very recently a heuristic
interpolation formula at $\beta=1$ incorporating exactly the limiting cases
of strong and weak absorption was suggested \cite{Fyodorov2004ii,Kuhl2005}
to describe reasonably well the available data at moderate absorption.
Therefore, an exact analytical treatment of the $\beta=1$ case remains a
considerable theoretical challenge.

In this Letter we present a novel approach to the problem which allows us to
derive the joint distribution function of the real and imaginary parts of
the local Green function exactly at arbitrary absorption for the whole
regime of the gradual TRS breaking between the $\beta=1$ and $\beta=2$
symmetry classes.

The Hamiltonian $\hat{H}$ of the chaotic system gives rise to $N$ levels
(eigenfrequencies) characterized locally in the relevant range of the energy
$E$ by the mean level spacing $\Delta$. We consider, as usual,
dimensionless quantities expressed in units of $\Delta$,
$K(E)\equiv(N\Delta/\pi)G(E)$, and define the distribution of interest as
follows:
\begin{equation}\label{P(u,v)} \mathcal{P}(u,v) =
\left\langle\delta(u-\mathrm{Re\,}K)\,\delta(v+\mathrm{Im\,}K)\right\rangle\,.
\end{equation}
Angular brackets stand for the ensemble averaging. In such units the mean
LDoS $\langle{v}\rangle=1$. The function $iK=Z$ has also the meaning of the
normalized cavity impedance $Z$ \cite{Hemmady2005}.

We start with establishing the general relation between the joint
distribution function (\ref{P(u,v)}) at \emph{finite} absorption (assumed to
be uniform, $\Gamma>0$) and the energy autocorrelation function
\begin{eqnarray}\label{C_omega}
C_{\Omega}(z_-,z_+) &=& \left\langle
\frac{1}{z_{-}-i0-K_{0}(E-\Omega/2-i0)}\right.
\nonumber \\
&& \times \left.\frac{1}{z_{+}+i0-K_{0}(E+\Omega/2+i0)} \right\rangle
\end{eqnarray}
of the \emph{resolvents} of the local Green function $K_0$ at \emph{zero}
absorption ($\Gamma=0$). Distribution (\ref{P(u,v)}) can be obtained from
(\ref{C_omega}) by analytic continuation in $\Omega$ from a real to purely
imaginary value $\Omega=i\Gamma$ as follows. $K_0(E)$ is an analytic
function of the energy in the upper or lower half-plane and can be thus
analytically continued to the complex values: $K_0(E\pm i\Gamma/2)\equiv
u\mp iv$, $v>0$. This allows us to continue then analytically the
correlation function (\ref{C_omega}) from a pair of its real arguments to
the complex conjugate one: $z_+ = (z_-)^*\equiv z'+iz''$, $z''>0$. As a
result, function (\ref{C_omega}) acquires at $\Omega=i\Gamma$ the following
form:
\begin{eqnarray}\label{C}
C(z',z'') &\equiv& C_{\Omega=i\Gamma}(z_-,z_+) =
    \left\langle\frac{1}{(z'-u)^2+(z''+v)^2}\right\rangle \nonumber\\
&=& \int_{-\infty}^{\infty}\!\!du\!\int_{0}^{\infty}\!\!dv\,
    \frac{\mathcal{P}(u,v)}{(z'-u)^2+(z''+v)^2} \,.
\end{eqnarray}
The second line here is due to definition (\ref{P(u,v)}). To solve this
equation for $\mathcal{P}(u,v)$, we perform first the Fourier transform (FT)
$\widehat{C}(k,z''){\equiv}\int_{-\infty}^{\infty}\!dz'e^{ikz'}C(z',z'')$
with respect to $z'$ that leads to
\begin{equation}\label{Chat} \widehat{C}(k,z'') =
\int_{0}^{\infty}\!\!dv\,\widehat{\mathcal{P}}(k,v)\,
\frac{\pi\,e^{-|k|(z''+v)}}{z''+v}\,,
\end{equation}
where $\widehat{\mathcal{P}}(k,v)$ is the corresponding FT of
$\mathcal{P}(u,v)$. Being derived at $z''>0$, Eq.~(\ref{Chat}) can be
analytically continued to the whole complex $z''$ plane with a cut along
negative $\mathrm{Re\,}z''$. Calculating then the jump of
$\widehat{C}(k,z'')$ on the discontinuity line $z''=-v$ ($v>0$), we finally
get the following expression
\begin{equation}\label{Phat}
\widehat{\mathcal{P}}(k,v)
=\frac{\Theta(v)}{2\pi^2i}[\widehat{C}(k,-v-i0)-\widehat{C}(k,-v+i0)]\,,
\end{equation}
with the Heaviside step function $\Theta(v)$. The inverse FT of
(\ref{Phat}) yields the desired distribution.

This relationship is one of our central results. It resembles (and reduces
to) the well-known relation between the spectral density of states and the
imaginary part of the corresponding resolvent operator when the case of one
real variable is considered. In contrast, the case of the distribution
function of two real variables requires to deal with the two-point
correlation function. Physically, the latter is a generalized susceptibility
which describes a response of the system that allows one to treat
(\ref{Phat}) in the sense of a fluctuation dissipation relation: The l.h.s.
there stands for the \emph{distribution} (of $K$) in the presence of
dissipation / absorption whereas the \emph{correlator} (of resolvents of $K$)
in the r.h.s. accounts for fluctuations in the system, i.e. for arbitrary
order
correlations in the absence of absorption.

We proceed now with applications. The main advantage of the derived relation
is that the correlator is a much easier object to calculate analytically
than the distribution and such a calculation for ideal systems at zero
absorption has actually already been performed in many interesting cases.
Let us consider the chaotic cavity mentioned already. In this case an exact
result for the correlation function (\ref{C_omega}) has been previously
obtained by us in Refs.~\cite{Fyodorov1997,Fyodorov1997i}. Its analytic
continuation to complex $\Omega=i\Gamma$ can be represented generally as
follows:
\begin{equation}\label{Cgeneric}
C(z',z'') = \frac{1}{z'^2+(z''+1)^2} + \frac{1}{4}\left(
\frac{\partial^2}{\partial z'^2} \!+\! \frac{\partial^2}{\partial z''^2}
\right) \mathcal{F}(\tilde{x})\,,
\end{equation}
where it is important that the function $\mathcal{F}(\tilde{x})$ depends on
$z'$ and $z''$ only via the scaling variable
$\tilde{x}\equiv(z'^2+z''^2+1)/2z''>1$. Its explicit form depends on the
symmetry present (e.g. preserved or broken TRS), the following common
structure being however generic:
\begin{widetext} 
\begin{equation}\label{Fgeneric}
\mathcal{F}(\tilde{x}) =
\int_{-1}^{1}\!d\lambda_0\!\int_{1}^{\infty}\!d\lambda_1\int_{1}^{\infty}\!d\lambda_2
\,f(\{\lambda\})\,e^{-\gamma(\lambda_1\lambda_2-\lambda_0)/2}
\left[ \frac{(\tilde{x}+\lambda_0)^2}{(\tilde{x}+\lambda_1\lambda_2)^2-
(\lambda_1^2-1)(\lambda_2^2-1)}\right]^{1/2}\,.
\end{equation}
Here, the dimensionless parameter $\gamma\equiv2\pi\Gamma/\Delta$ (i.e.
absorption width $\Gamma$ in units of the mean level spacing $\Delta$)
accounts for the absorption strength. The \emph{real} function
$f(\{\lambda\})$ is the only symmetry dependent term. In the crossover
regime of gradually broken TRS it can be represented explicitly as follows
\cite{Fyodorov1997i}:
\begin{equation}\label{f}
f(\{\lambda\}) = \left\{ (1-\lambda_0^2)(1+e^{-2Y})
-(\lambda_1^2-\lambda_2^2)(1-e^{-2Y}) + 4y^2\mathcal{R}[
(1-\lambda_0^2)e^{-2Y} + \lambda_2^2(1-e^{-2Y})] \right\}
\frac{e^{-2y^2(\lambda_2^2-1)}}{\mathcal{R}^2}\,,
\end{equation}
\end{widetext} 
with
$\mathcal{R}=\lambda_0^2+\lambda_1^2+\lambda_2^2-2\lambda_0\lambda_1\lambda_2-1$
and $Y\equiv{y^2}(1-\lambda_0^2)$, where $y$ denotes a crossover driving
parameter. Physically, $y^2\sim{\delta}E_y/\Delta$ is determined by the
energy shift $\delta{E}_y$ of energy levels due to a TRS breaking
perturbation (e.g., weak external magnetic field in the case of quantum
dots). Such an effect is conventionally modelled within the framework of RMT
by means of the ``Pandey-Mehta'' Hamiltonian \cite{Pandey1983},
$\hat{H}=\hat{H}_{S}+i(y/\sqrt{N})\hat{H}_{A}$, with $\hat{H}_{S}$
($\hat{H}_{A}$) being a random real symmetric (antisymmetric) matrix with
independent Gaussian distributed entries. The limit $y\to0$ or $\infty$
corresponds to fully preserved or broken TRS, respectively.

Now we apply relation (\ref{Phat}) to Eq.~(\ref{Cgeneric}) and then perform
the inverse FT to get $\mathcal{P}(u,v)$. Relegating all technical details
to a more extended publication, we emphasize the most important points. The
nontrivial contribution to the distribution comes from the second
(``connected'') part of the correlation function (\ref{Cgeneric}) whereas
the first (``disconnected'') one is easily found to yield the singular
contribution $\delta(u)\delta(v-1)$. A careful analysis shows that due to
specific $\tilde{x}$-dependence given by Eq.~(\ref{Fgeneric}) the above
described procedure for the analytic continuation of the connected part of
$\widehat{C}(k,z'')$ is equivalent to continuing $\mathcal{F}(\tilde{x})$
analytically and taking the jump at $\tilde{x}=-x\pm{i}0$, with
$$
x\equiv\frac{u^2+v^2+1}{2v}>1\,.
$$
The nonzero imaginary part $F(x)=\mathrm{Im\,}\mathcal{F}(-x)$ is thus
determined at given $x$ by the integration region
$\mathcal{B}_x=\{(\lambda_1,\lambda_2)\,|$ $1\leq\lambda_{1,2}<\infty,$
$(\lambda_1\lambda_2-x)^2<(\lambda_1^2-1)(\lambda_2^2-1)\}$, where the
square root in (\ref{Fgeneric}) attains pure imaginary values. Taking into
account the identity $\partial^2F(x)/\partial u^2+\partial^2F(x)/\partial
v^2 = v^{-2}\frac{d}{dx}(x^2-1)\frac{d}{dx}F(x)$ valid for $x^2\neq1$, we
arrive finally at
\begin{equation}\label{Pgeneric}
\mathcal{P}(u,v) = \frac{1}{4\pi^2v^2}\frac{d}{dx}(x^2-1)\frac{dF(x)}{dx}
\equiv \frac{1}{2\pi v^2}P_0(x)\,.
\end{equation}
This distribution is easily checked to be invariant under the change
$iK\to1/iK$, meaning physically that the impedance and its inverse have one
and the same distribution function.

Such a form of the distribution is completely generic, as all the symmetry
specific dependencies were not essential for the above discussion. It can be
shown \cite{Fyodorov2004ii} by exploiting the well-known relation $S =
\frac{1-i\kappa K}{1+i\kappa K} \equiv \sqrt{r}e^{i\theta}$ between the
scattering matrix $S$ and the local Green function (known as $K$ function in
this context), that the representation given by the second equality in
Eq.~(\ref{Pgeneric}) is a consequence of the two following properties at the
so-called \emph{perfect coupling}, $\kappa=1$  \cite{coupling}: (i) the
uniform distribution of the scattering phase $\theta\in(0,2\pi)$; and (ii)
the statistical independence of $\theta$ and the $S$ matrix modulus. This
establishes also a physical meaning of $x$ by relating it to the
\emph{reflection coefficient} $r$, thus $P_0(x)$ being the normalized
distribution of $x=\frac{1+r}{1-r}$. Remarkably, Eq.~(\ref{Pgeneric}) relates
the distribution of the local Green function in the closed system to that of
reflection in the perfectly open one.

Both these properties can be verified using the methods
of Ref.~\cite{Brouwer1997ii} but only in the cases of preserved or
completely broken TRS. Our approach proves it generally for the crossover
regime at an arbitrary degree of TRS breaking. Taking into account our
findings, we can bring the final result to the following form:
\begin{eqnarray}\label{Fcross}
F(x) &=&
\!\int_{-1}^{1}\!d\lambda_0\!\int\!\!\int_{\mathcal{B}_x}\!d\lambda_1d\lambda_2
\,f(\{\lambda\})\,e^{-\gamma(\lambda_1\lambda_2-\lambda_0)/2} \nonumber\\ &&
\times \frac{ (x-\lambda_0) }{
[(\lambda_1^2-1)(\lambda_2^2-1)-(\lambda_1\lambda_2-x)^2]^{1/2}}\,.
\end{eqnarray}
At arbitrary values of the crossover parameter $y$ the obtained result can
be treated only numerically. Further analytical study is possible in the
limiting cases of pure symmetries considered below.

The simplest case of unitary symmetry ($\beta=2$) \cite{Beenakker2001} is
correctly reproduced from (\ref{Fcross}) as $y\to\infty$. We find
\cite{GUE}:
\begin{equation}\label{P(x)}
P_0(x) = \frac{{\mathcal N}_{\beta}}{2}
\left[A\,\left(\alpha(x+1)/2\right)^{\beta/2}+B\right]e^{-\alpha(x+1)/2}\,,
\end{equation}
where it is convenient for the subsequent use to introduce the absorption
parameter $\alpha\equiv\gamma\beta/2$ scaled with the symmetry index
$\beta$, $\alpha$-dependent constants being $A\equiv e^{\alpha}-1$ and
$B\equiv1+\alpha-e^{\alpha}$, and the normalization constant
$\mathcal{N}_2=1$.

As to the case of orthogonal symmetry ($\beta=1$),  no general result was
available in the literature. Expression (\ref{P(x)}) (with
$\mathcal{N}_{\beta}=\alpha/(A\Gamma(\beta/2{+}1,\alpha)+Be^{-\alpha})$ and
$\Gamma(\nu,\alpha)=\int_{\alpha}^{\infty}\!\!dt\,t^{\nu-1}e^{-t}$) was
suggested in Ref.~\cite{Fyodorov2004ii} (see also \cite{Kuhl2005}) to be an
appropriate interpolation formula at $\beta=1$. It incorporates correctly
both known limiting cases of weak or strong absorption and a reasonable
agreement with available numerical and experimental data was found  in a
broad range of the absorption strength. We proceed with providing an exact
analytical treatment of this case which amounts to investigating
(\ref{f}) and (\ref{Fcross}) at $y=0$. Fortunately, further simplifications
are
possible if one considers the integrated probability distribution
\begin{equation}\label{W(x)}
W(x) \equiv -\frac{x^2-1}{2\pi}\frac{dF(x)}{dx} =
\int_x^{\infty}dx\,P_0(x)\,,
\end{equation}
which is a positive monotonically decaying function by definition. To this
end, we note that it is useful to switch to the parametrization of
Ref.~\cite{Verbaarschot1985} to carry out the threefold integration. The
latter turns out to yield then a sum of decoupled terms and, after some
algebra, we have been able to cast the result in the following final form:
\begin{eqnarray}\label{Wgoe}
W(x) &=& \frac{x+1}{4\pi}\Bigl[f_1(w)g_2(w)+f_2(w)g_1(w)\nonumber\\
&& +h_1(w)j_2(w)+h_2(w)j_1(w)\Bigr]_{w=(x-1)/2}\,,\quad
\end{eqnarray}
with auxiliary functions defined as follows:
\begin{eqnarray}
f_1(w) &=& \int_w^{\infty}\!\!dt
\frac{\sqrt{t|t-w|}\,e^{-\gamma{t}/2}}{(1+t)^{3/2}} [1-e^{-\gamma}+t^{-1}]\,,
\nonumber\\
g_1(w) &=& \int_w^{\infty}\!\!dt \frac{1}{\sqrt{t|t-w|}} \frac{e^{-\gamma
t/2}}{(1+t)^{3/2}}\,,
\nonumber\\
h_1(w) &=& \int_w^{\infty}\!\!dt \frac{\sqrt{|t-w|}\,e^{-\gamma
t/2}}{\sqrt{t(1+t)}} [\gamma+(1-e^{-\gamma})(\gamma t-2)]\,,
\nonumber\\
j_1(w) &=& \int_w^{\infty}\!\!dt \frac{1}{\sqrt{t|t-w|}} \frac{e^{-\gamma
t/2}}{\sqrt{1+t}}\,,
\end{eqnarray}
their counterpart with the index 2 being given by the same expression save
for the integration region $t\in[0,w]$ instead of $[w,\infty)$. For an
illustration of our findings the reflection distribution
$P(r)=\frac{2}{(1-r)^2}P_0(\frac{1+r}{1-r})$ is shown on Fig.~1.

\begin{figure}[t]
\includegraphics[width=0.47\textwidth]{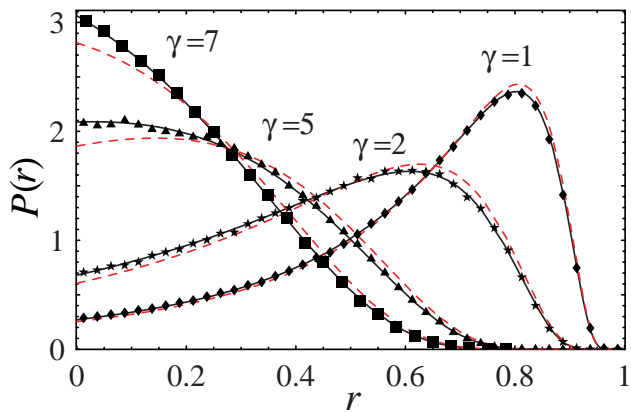}
\caption{The reflection coefficient distribution in chaotic systems
invariant under time-reversal at perfect coupling is plotted at moderate
values of the absorption strength $\gamma=2\pi\Gamma/\Delta$, where
systematic deviations between the exact result drawn from Eq.~(\ref{Wgoe})
(solid lines) and the interpolation expression (\ref{P(x)}) at $\beta=1$
(dashed lines) are most noticeable. Symbols stand for numerics done for
$10^3$ realizations of 500$\times$500 random GOE matrices.}
\end{figure}

Finally, we mention that the general case of arbitrary transmission $T<1$
can be mapped \cite{Brouwer1995,Savin2001} onto that of perfect one
considered so far. The scattering phase $\theta$ is then no longer uniformly
distributed and gets statistically correlated with $x$. However, their joint
distribution $P(x,\theta)$ can be found \cite{Fyodorov2004ii} to be again
determined by $P_0(x)$ as follows:
\begin{equation}\label{P(x,theta)}
P(x,\theta) = \frac{1}{2\pi}
P_0(xg-\sqrt{(x^2-1)(g^2-1)}\cos\theta) \,,
\end{equation}
with $g\equiv2/T-1$. This equation provides us further with distributions of
the phase and reflection coefficient which were recently studied
experimentally \cite{Kuhl2005}.

In summary, we provided a new approach to statistical properties of the
local Green function in quantum chaotic or disordered absorptive systems of
any symmetry class. It would be highly interesting to extend current
experimental studies \cite{Kim2005} of the crossover regime to check our
findings. Although the validity of explicit formulas given above is
restricted to the completely ergodic (``zero-dimensional'') case, it is
actually possible to adopt the above method \cite{prog} to quasi-one (or
higher) dimensional situations when Anderson localization effects play
already an important role \cite{Mirlin2000}.

We thank V.V. Sokolov for useful comments. The financial support by the
SFB/TR 12 der DFG (D.V.S. and H.-J.S.) and EPSRC grant EP/C515056/1 ``Random
Matrices and Polynomials'' (Y.V.F.) is acknowledged.


\begin{thebibliography}{25}
\bibitem{GUE}
$4y^2\!\int_{1}^{\infty}d\lambda_2\lambda_2^2e^{-2y^2(\lambda_2^2-1)}\to1$ as
$y\to\infty$, so that one can put $\lambda_2{=}1$ everywhere else in
(\ref{Fcross}). As a result, the $\lambda_1$ integration in $F(x)$ shrinks
to a
point $\propto\delta(\lambda_1-x)$, then the integration over $\lambda_0$ is
easily performed to yield Eq.~(\ref{P(x)}).
\end{thebibliography}
%

\end{document}